\documentclass[]{spie}  

\usepackage{amsmath,amsfonts,amssymb}
\usepackage{graphicx}
\usepackage[colorlinks=true, allcolors=blue]{hyperref}
\usepackage{subcaption}

\title{Exploring calibration algorithms to maximize the null depth in KPIC’s vortex fiber nulling mode}

\author[a]{Sofia Hillman}
\author[b]{Daniel Echeverri}
\author[a]{Maxwell A. Millar-Blanchaer}
\author[b]{Jerry Xuan}
\author[b]{Garreth Ruane}
\author[b]{Dimitri Mawet}

\affil[a]{Department of Physics, University of California, Santa Barbara, Broida Hall, Santa Barbara, CA 93106-9530, USA}
\affil[b]{California Institute of Technology, 1200 E. California Blvd, Pasadena, CA 91125, USA}

\authorinfo{Further author information: (Send correspondence to Sofia Hillman)\\S. Hillman: E-mail: hillman.sofia@gmail.com\\  M. Millar-Blanchaer: E-mail: maxmb@ucsb.edu}

\begin{document} 
\maketitle

\begin{abstract}
Vortex fiber nulling (VFN) is a new interferometric technique with the potential to unlock the ability to detect and spectroscopically characterize exoplanets at angular separations smaller than the conventional diffraction limit of $\lambda$/D. In early 2022, a VFN mode was added to the Keck Planet Imager and Characterizer (KPIC) instrument suite on Keck II. VFN operates by adding an azimuthal phase ramp to the incident wavefront so that light from the star at the center of the field is prevented from coupling into a single-mode fiber. One of the key performance goals of VFN is to minimize the ratio of on-axis starlight coupling to off-axis planet coupling, which requires minimizing the wavefront aberrations of light being injected into the fiber. Non-common path aberrations can be calibrated during the daytime and compensated for with the KPIC deformable mirror during nighttime observing. By applying different amplitudes of low-order Zernike modes, we determine which combinations maximize the system performance. Here we present our work developing and testing different procedures to estimate the incident aberrations, both in simulation and on the Keck bench. The current iteration of this calibration algorithm has been used successfully for VFN observing, and there are several avenues for improvement.
\end{abstract}


\section{INTRODUCTION}
\label{sec:intro}  

The Keck Planet Imager and Characterizer (KPIC) is an astronomical instrument installed at Keck II on Maunakea, Hawaii. High contrast imaging and high resolution spectroscopy are key capabilities which allow KPIC to be successful for directly imaging exoplanets \cite{KPIC2016}. VFN is an interferometric method for suppressing starlight. It has the potential to allow detection of exoplanets at smaller angular separations than the traditional diffraction limit of $\lambda$/D, where $\lambda$ is the operating wavelength and D is the telescope diameter. The design and planned implementation for a vortex fiber nulling (VFN) mode was described in a 2016 paper \cite{KPIC2016} and the VFN mode was adding to KPIC in early 2022. The first light and performance are described in an SPIE paper authored by Daniel Echeverri which is also in these proceedings\cite{ECHEVERRI2023}. VFN operates by utilizing a mask in the pupil plane with an azimuthally increasing phase ramp pattern to shift the phase of incident light and is described in a paper by Garreth Ruane from 2016\cite{KPIC2016} and was demonstrated experimentally with monochromatic light as described in a 2019 paper \cite{ECHEVERRI2019}. Light is then passed to a single mode fiber (SMF), and coupling into the fiber is defined by the overlap integral, given in Equation \ref{eq:overlap}. Thus, placing the vortex mask on axis with the host star nulls much of the starlight while allowing light from off axis sources, such as exoplanets, to be detected\cite{KPICVFN2019}.

\subsection{Principles of VFN}

The first application of interferometric stellar nulling for exoplanet detection being introduced in 1978\cite{BRACEWELL1978}. VFN specifically is a relatively new concept and was first proposed in 2018\cite{KPICVFN2019}. Since then, technological advancements have been made and VFN technology, along with the next generation of telescopes, may allow for observation of rocky exoplanets in the habitable regions of their host stars \cite{VFN2019}. The principle behind VFN is placing a vortex mask with an azimuthal phase ramp in front of a SMF. The mask imparts a phase to incident electric fields depending on the azimuthal coordinate of incidence. Coupling of light which passes through the mask into the fiber is determined by the overlap integral\cite{VFN2019} which has the form of equation \ref{eq:overlap}. Here, $\mathbf{r}$ is the polar coordinate which defines a location on the face of the SMF, $E(\mathbf{r})$ is the electric field incident on the vortex mask, and $\Psi(\mathbf{r})$ is the fundamental mode of the fiber.

\begin{equation} \label{eq:overlap}
    \eta = \left| \int E(\mathbf{r})\Psi(\mathbf{r})d\mathbf{r} \right|^2
\end{equation}

\begin{figure}
\centering
  \begin{subfigure}[t]{.42\textwidth}
  \centering
    \includegraphics[width=.9\textwidth]{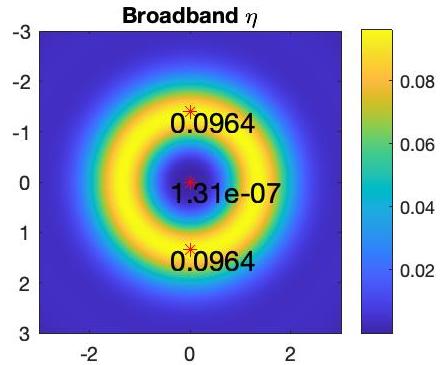}
    \caption{\label{fig:goodcouplingmaps1}No aberrations}
  \end{subfigure}
    \begin{subfigure}[t]{.4\textwidth}
  \centering
    \includegraphics[width=.9\textwidth]{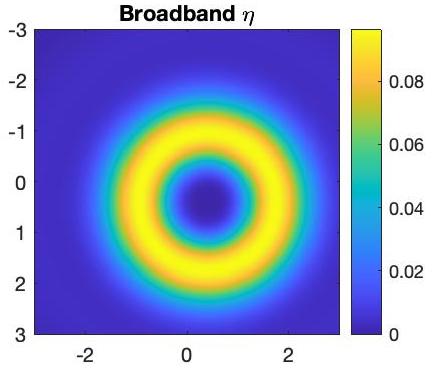}
    \caption{\label{fig:goodcouplingmaps2}0.1 waves RMS Zernikes 2 \& 3 (tip/tilt)}
  \end{subfigure}
  \begin{subfigure}[t]{.42\textwidth}
  \centering
    \includegraphics[width=.9\textwidth]{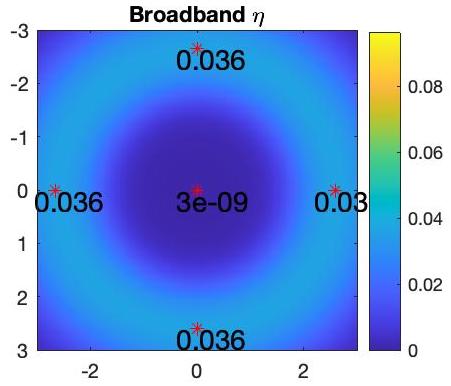}
    \caption{\label{fig:goodcouplingmaps3}0.3 waves RMS Zernike 4 (defocus)}
  \end{subfigure}
  \begin{subfigure}[t]{.42\textwidth}
  \centering
    \includegraphics[width=.9\textwidth]{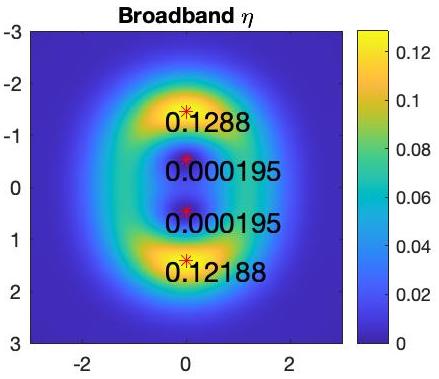}
    \caption{\label{fig:goodcouplingmaps4}0.05 waves RMS Zernike 5 (astigmatism)}
  \end{subfigure}
\caption{\label{fig:goodcouplingmaps}Simulated coupling efficiency maps with four different aberrations. $x$ and $y$ units are arbitrary spatial units on the face of the vortex mask.}
\end{figure}

The fundamental mode of a SMF is azimuthally symmetric \cite{VFN2019}. Point sources centered on the SMF will produce electric fields with antisymmetry to the fundamental mode of the SMF. Thus, this integral will result in 0 coupling and so such sources will not couple into the fiber. However, off-axis sources will not produce electric fields which are symmetric about the  axis of the vortex. Without the nominal symmetry, the overlap integral will yield a value greater than zero, indicating that some light from these sources will couple into the fiber. This concept can be applied to imaging exoplanets by aligning the host star on the axis of the vortex mask. Then, light from the star will not couple (will be nulled) and light from the off-axis exoplanet will at least partially couple. Null depth is defined as the coupling of star after it has been nulled. Coupling is defined as the fraction of light from a source that enters the fiber. Thus, the goal of a VFN mask is to minimize stellar coupling ($\eta_S$) and maximize off-axis coupling ($\eta_P$). Figure \ref{fig:goodcouplingmaps1} depicts an ideal coupling map for charge 2 VFN. The donut shape is evident since axial light is minimally coupled into the fiber while off-axis light couples. The other sub-figures in Figure \ref{fig:goodcouplingmaps} will be discussed in Section \ref{sec:simulations}.

\begin{figure}
\centering
\includegraphics[width=0.45\textwidth]{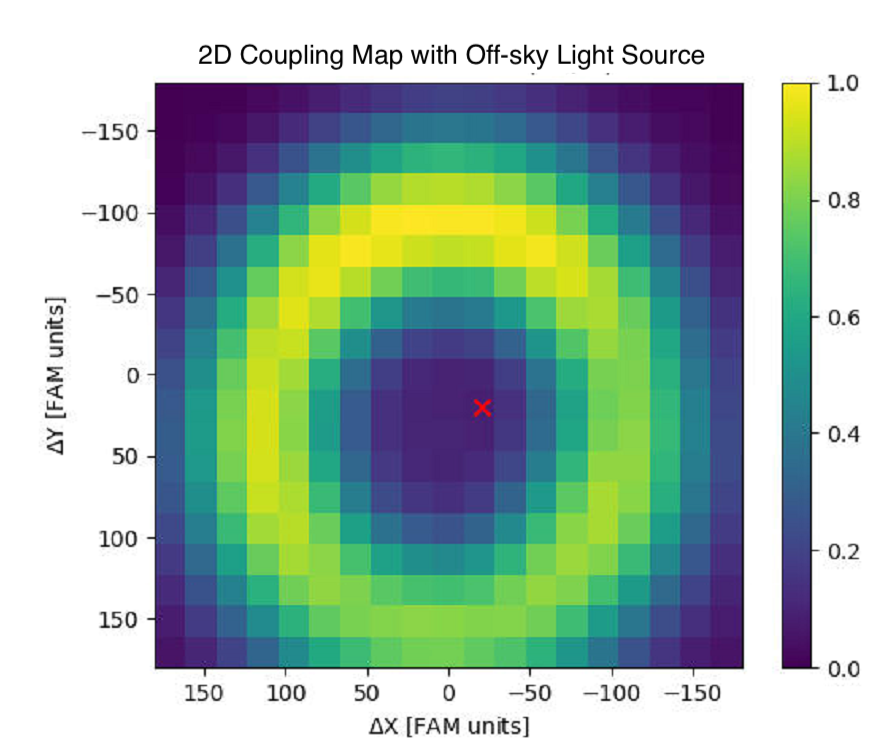}
\includegraphics[width=0.45\textwidth]{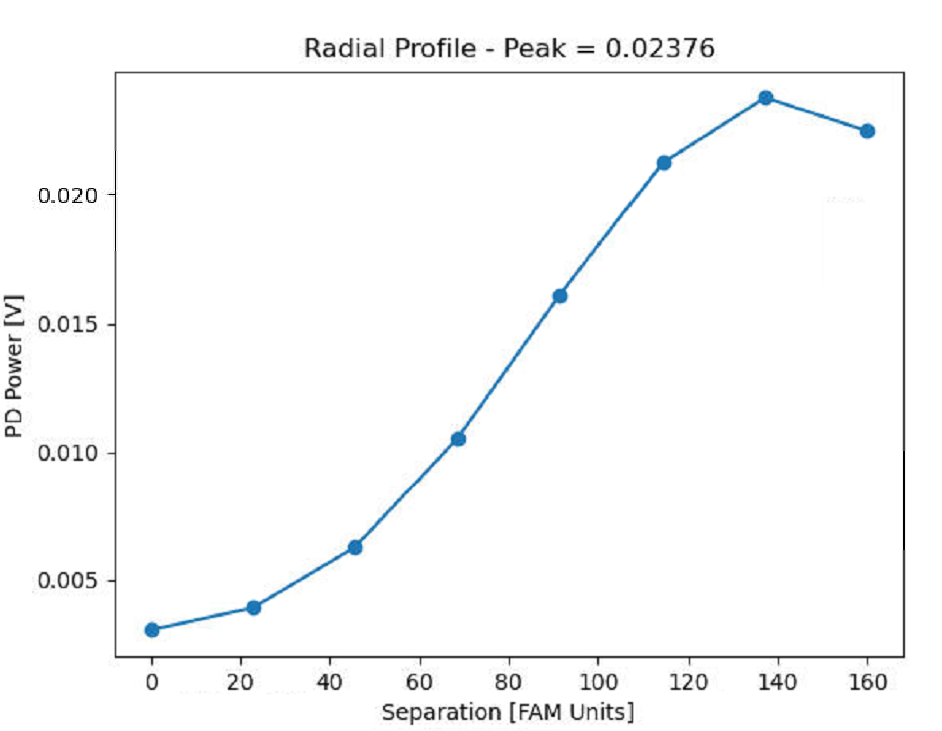}
\caption{\label{fig:radialProfile} Example VFN coupling map and average radial profile of the coupling map. $x$ and $y$ units (FAM units) are arbitrary spatial units in the focal plane which are proportional to the angular separation of a point from the null. The colorbar notes the percent coupling into the SMF. PD Power is the measure of the voltage passed from the SMF to the photodiode. Refer to Section \ref{aberration} for a description of how these measurements are made.}
\end{figure}

Figure \ref{fig:radialProfile} depicts an example coupling map which was generated off-sky with the KPIC VFN optics and internal calibration light sources to simulate a point source. The first plot shows the ``donut''-like shape of a VFN coupling map and the second plot shows the average radial profile over this coupling map. The units of photodiode (PD) power represent the signal which couples into the SMF at different angular separations from the null. These plots show coupling performance before calibrations were done so it gives a sense of the typical starting conditions. The donut shape is not perfectly symmetric and the null (indicated by the red \textbf{X}) is not centered in the donut shape. 

Given theoretical performance estimates made by the KPIC VFN team \cite{KPICVFN2019}, the KPIC VFN mode was designed with a charge 2 vortex in the pupil plane of KPIC which is specialized for K-band light. With a smaller inner working angle than traditional coronagraphs, the KPIC VFN mode will be able to directly image exoplanets at smaller angular separations than conventional coronagraphs. With future large telescopes, VFN may facilitate observation and characterization of rocky planets in the habitable zones of their host stars. The habitable zones of many host stars are inside the inner working angle of traditional coronagraphs. Further, planets closer to their stars have a higher reflected light flux, which will allow easier spectroscopy.

\subsection{Wavefront Aberrations}\label{aberration}

Removing wavefront aberrations is especially important for VFN because stellar nulling relies on the overlap integral yielding 0 at the center of the SMF, which requires a flat wavefront. Aberrations can cause both increased axial coupling or decreased off-axis coupling. Thus, the system is sensitive to aberrations including adaptive optics (AO) residuals. However, the focus of this project is on correcting non-common path aberrations (NCPAs). NCPAs are asymmetries in the electric field which are introduced downstream from the wavefront sensor and so they must be removed by instrument calibration rather than by AO. In this project, we use the Zernike polynomials, specifically the Noll expansion, to characterize aberrations. The goal of this project is to determine an efficient and effective method to identify existing NCPAs so that they can be accounted for and corrected during the VFN daytime calibration procedure.

Detecting and correcting aberrations is done using a single photodiode to measure light which has coupled into the SMF after passing through the vortex mask. By scanning the calibration source across the focal plane and using a tip/tilt mirror to measure coupling maps along with the calibration procedure described here, the existing Zernike aberrations can be estimated. Then, the opposite Zernike aberration can be applied to the deformable mirror to correct the existing aberration. The difficulty in doing this is using the single photodiode to understand the two-dimensional aberrations that exist. The rest of this paper explores methods to most accurately yet efficiently identify aberrations.

KPIC VFN has been successfully operating since 2022 when the optics were installed. The calibration procedure that has previously been used can mitigate NCPAs enough that the null depth is limited by on-sky atmospheric wavefront residuals rather than by the wavefront residuals of the instrument. However, there are two primary downfalls of this procedure. First, it requires performing multiple 2-dimensional scans of the coupling into the SMF across the face of the vortex mask. Thus, the whole procedure takes 2-3 hours. Second, due to time constraints, this procedure solely optimizes the null depth at the cost of off-axis symmetry. Thus, calibrating in this way is sensitive to planets in specific regions around the star and insensitive to those in others. The new procedure described here attempts to improve calibration run-time and off-axis symmetry. Considering off-axis symmetry may decrease null depth slightly but not so much that null depth is limited by NCPAs. Additionally, it comes with the benefit of off-axis symmetry.

\section{EXPLORING CALIBRATION METHODS}\label{sec:simulations}

As discussed, the key goals of VFN are to maximize null depth and off-axis throughput. Ultimately, the goal of this project is to minimize the ratio $\eta_S/\eta_P^2$. With KPIC, on-sky observers have found that this ratio is limited to around $1/7$ by AO residuals. For the charge 2 vortex installed at KPIC, the maximum off-axis throughput generally occurs at 60 milliarcsecond (mas) separation from the null so this is where $\eta_P$ is measured for this project. This section summarizes the simulations done to test calibration approaches for each Zernike. The calibration procedure attempts to maximize null depth and off-axis throughput by tuning the Zernikes that affect these coupling properties.

\begin{figure}
\centering
\includegraphics[width=1\textwidth]{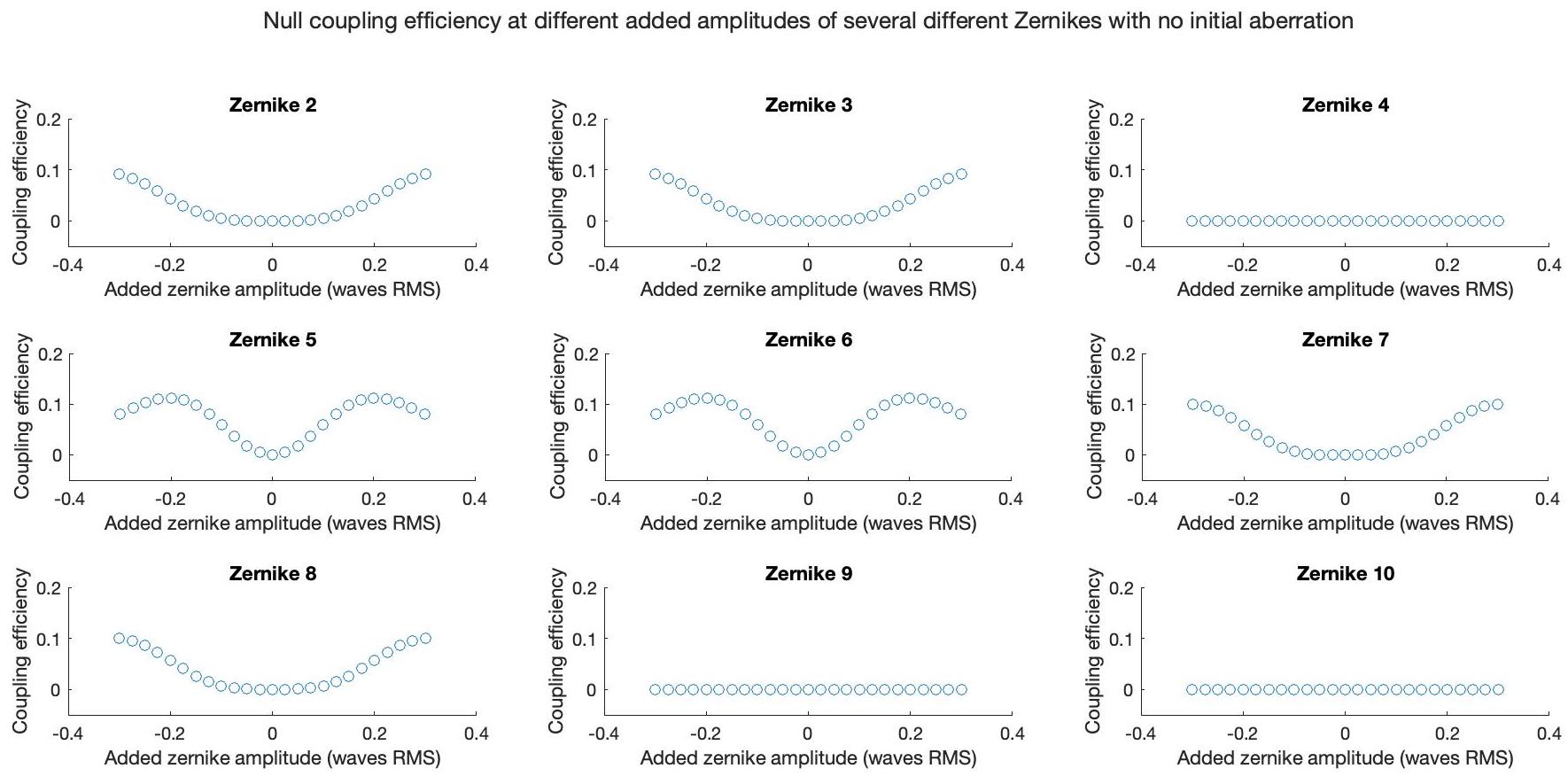}
\caption{\label{fig:noAbberation}Simulated coupling efficiency at the null with added Zernike aberrations of amplitudes ranging from -0.3 waves RMS to 0.3 waves RMS. Coupling efficiency is the percent of light from a point source which couples into the SMF. Waves RMS is calculated using this formula: waves RMS $=\sqrt{\overline{\phi^2}-\overline{\phi}^2}$ where $\phi$ is the wavefront error of an incident wave.}
\end{figure}

Figure \ref{fig:noAbberation} depicts the simulated coupling efficiency at the null that occurs when different individual aberrations are present. Starting with a model flat map (no aberrations) the null coupling was measured after adding different amplitudes of different Zernikes, one at a time. In the figure, it is clear that there is no initial aberration because the null depth is maximized when no aberration is added. The case where there are NCPAs present can be simulated by adding an initial aberration, then going through the same process. For these simulations, the amplitude range is -0.3 waves RMS to 0.3 waves RMS with a step size of 0.025 waves RMS. In practice with KPIC, NCPAs present are generally encompassed by this range and the step size provides enough sensitivity that aberrations smaller than the step size are smaller than typical noise.

There are a few important takeaways from Figure \ref{fig:noAbberation}, which are also theoretically expected \cite{ECHEVERRI2021}. First, null depth is maximized when there are no aberrations, which is clear since the minima of each subplot is when no Zernike aberration is added. Next, the null is more sensitive to aberrations in certain Zernikes than aberrations in others. Specifically, aberrations in Zernikes 5 and 6 (first order astigmatism) clearly have a stronger effect on null depth than the other Zernikes. Because of this, astigmatism will be a major focus of this project. Conversely, Zernikes 2 and 3 (tip/tilt) and Zernikes 7 and 8 (first order coma) have a weaker impact on null depth. Finally, Zernike 4 (defocus) and Zernikes 9 and 10 (first order trefoil) have almost no affect on null depth. Notably, these Zernikes do impact off-axis coupling and so they cannot be ignored in calibrations.

\subsection{Tuning astigmatism}\label{sec:astigmatism}

As depicted in Figure \ref{fig:noAbberation}, astigmatism aberrations have the strongest effect on null depth so correcting them will be the primary focus of this project. However, an additional complication is that when astigmatism aberrations are present, there can be degeneracies in the individual Zernikes' effect on null depth.

Astigmatism aberrations, as well as coma aberrations, produce an effect called null splitting. These aberrations create multiple ``nulls", depicted in  Figure \ref{fig:goodcouplingmaps4}, which are off-axis of the vortex mask and are not as deep as a single on-axis null. The simulated perfect null has a depth of order $10^{-7}$ while the simulated split nulls only have depths on the order of $10^{-4}$, as shown in the labeled points. Because of this, tuning for the deepest null should correct null splitting caused by astigmatism.

\begin{figure}
\centering
\includegraphics[width=1\textwidth]{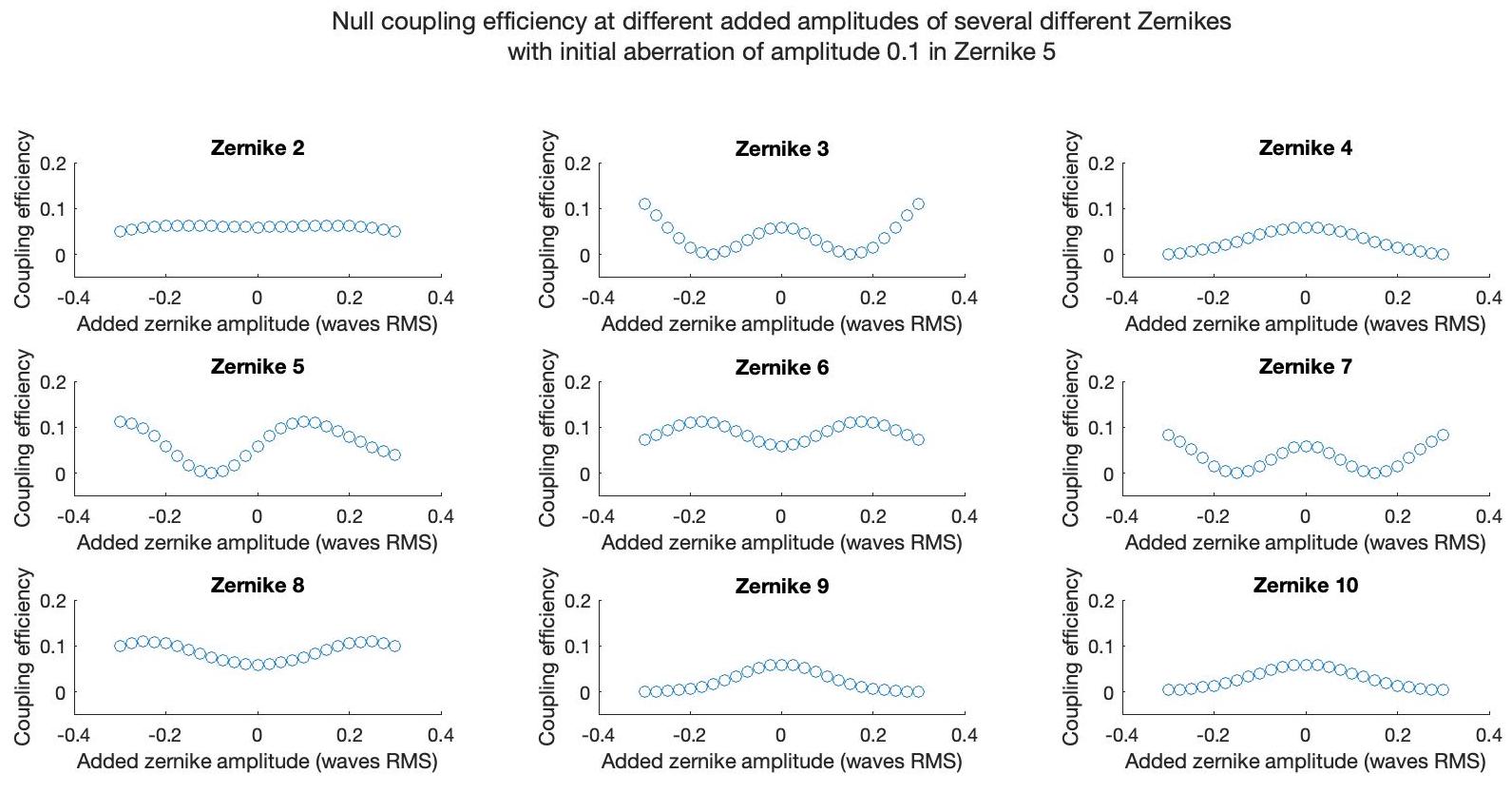}
\caption{\label{fig:zern5aberration}Simulated coupling efficiency at the null with dded Zernike aberrations of 
amplitudes ranging from -0.3 waves RMS to 0.3 waves RMS Zernike wavefront error. There is an initial aberration of amplitude 0.1 waves RMS initially present in Zernike 5. It is clear in the plot for Zernike 5 that the null coupling efficiency is minimized by adding -0.1 waves RMS to counterbalance the initial aberration. Note that when scanning the null, adjusting the amplitude of other Zernikes (which do not have initial aberrations) can also change the null depth.}
\end{figure}

Figure \ref{fig:zern5aberration} depicts the effect on null depth when different amplitudes of the first nine zernikes are added in different amplitudes to a map which already has an aberration of amplitude 0.1 waves RMS in Zernike 5. In the plot of Zernike 5, it is clear that the null coupling efficiency is minimized with an added amplitude of -0.1 waves RMS, as expected. Adding this amplitude effectively counterbalances the initially added aberration. Thus, this plot suggests that it is possible in a calibration procedure to find the initial NCPAs by adding different amplitudes of several Zernikes to determine which minimizes the null.

However, another notable effect shown in Figure \ref{fig:zern5aberration} is some degeneracy in null depth when astigmatism is present. Applying Zernike aberrations that are not opposite of existing aberrations may also lower the null depth. This is clear in the plots for Zernikes 3, 6, 7, and 8, which show coupling minima not at 0 waves RMS. Applying these ``corrections'' would reach a local minimum in null coupling but prevent observers from reaching the global minimum null coupling, which occurs when there are no aberrations present. The most obvious solution to this problem is to do a multidimensional calibration; rather than applying corrections in one Zernike at a time, this procedure would apply different combinations of Zernike corrections to determine which combination minimizes the null. However, calibrations need to be efficient and multidimensional scans increase the calibration time exponentially with the number of dimensions. As discussed previously, an typical amplitude range is -0.3 to 0.3 with step size 0.025, for a total of 25 amplitudes. Additionally, scans typically include Zernikes 2-10 as well as 2nd and 3rd order astigmatism terms for a total of 13 Zernikes. In order to measure null depth for every combination of these Zernike amplitudes, this totals to $25^{13}$ scans! Since each photodiode measurement takes a finite amount of time, this multidimensional calibration is impractical. Multidimensional scans may be necessary in some cases, but it is best to opt for a more efficient solution for most Zernike calibrations.

Note that, in Figure \ref{fig:zern5aberration}, the minimum coupling efficiency when applying different values of Zernike 5 is on the order of $10^{-12}$, while the minimum null coupling efficiency when changing any other Zernike is on the order of $10^{-4}$. Thus, when scanning through different Zernike aberrations, the behavior of the null depth should be dominated by real aberrations. Since astigmatism aberrations affect null depth most significantly, these Zernikes orders should be corrected first. Higher orders of astigmatism produce similar affects on null depth so calibration should include tuning orders 1-3 of astigmatism.

\subsection{Tuning tip/tilt}\label{sec:tiptilt}

Another important sensitivity is to tip/tilt aberrations, which are Zernikes 2 and 3. These aberrations can be thought of as translations of the VFN coupling map. Refer to Figure \ref{fig:goodcouplingmaps2} for an example of an otherwise perfect coupling map with 0.1 waves RMS of both Zernikes 2 and 3. Since there are no asymmetries in the coupling map, it is easy to see where the center of the map should be.

When measuring null depth as described in Section \ref{sec:astigmatism}, SMF coupling is measured from the center of the VFN mask and so it is important that the center of the SMF is aligned perfectly to the VFN mask. If it is not, tuning this off-center point for null depth will not produce a symmetric coupling map or nearly as deep of a null. Hence, calibrating tip/tilt must be done at the beginning of the calibration procedure. As determined in Section \ref{sec:astigmatism}, however, it is not possible to determine existing tip/tilt aberrations by scanning the null when there are existing astigmatism aberrations. Hence, to begin the calibration procedure, the coupling map must be roughly centered by eye. While this method is crude, it can be redone computationally after astigmatism has been corrected.

\subsection{Tuning defocus}\label{sec:defocus}

As shown in Figure \ref{fig:noAbberation}, defocus has minimal effect on null depth. Further, Figure \ref{fig:goodcouplingmaps3} depicts the effect of defocus on off-axis coupling. Not only does defocus spread the coupling map's radial maxima further from the null, it decreases their value. Since defocus primarily impacts off-axis coupling and has no affect on off-axis coupling symmetry, it should be scanned right after tip/tilt calibrations to ensure sufficient off-axis throughput for further scans.

With little impact on null coupling, defocus must be tuned by maximizing off-axis throughput. However, a single 2D scan over the area of the SMF generally take several minutes depending on the exact step size between scanned points. This is inefficient, especially because multiple amplitudes of defocus must be tested. One solution is to measure eight points at 60mas, each separated by $\pi/4$ rad. The separation 60mas is chosen because, in the absence of aberrations, the radial maxima are at 60mas for KPIC VFN. Scanning eight points is a compromise between scanning few enough points to be efficient but scanning enough points to reveal the symmetry in the most common Zernike modes that are scanned. Specifically, first order astigmatism polynomials have four azimuthal extrema each which align with the eight points aforementioned. Adjusting off-axis coupling is complicated since it is important to not only maximize off-axis coupling but create a symmetric coupling map. Since defocus is azimuthally symmetric, this is not a concern and defocus can be tuned by maximizing mean off-axis coupling at eight points at 60mas.

\subsection{Tuning coma}

As shown in Figure \ref{fig:noAbberation}, null depth is sensitive to coma aberrations. Once astigmatism aberrations are not present, coma aberrations should be detectable by finding the amplitudes of coma which minimize null coupling. Both 1st and 2nd order coma should be scanned and corrected.

\subsection{Tuning trefoil}

Trefoil is the final Zernike to be corrected. Figure \ref{fig:noAbberation} shows that null depth is relatively insensitive to trefoil. Hence, it should be tuned using the same method as was used for tuning defocus. Again, the goal will be to maximize mean off-axis coupling at eight azimuthally distributed points at 60mas.

\section{CALIBRATION METHOD}\label{sec:method}

The goal of this project is to determine the most efficient but effective procedure to correct time-stable NCPAs present the KPIC VFN system. This is important because observing with VFN requires optimizing the ratio $\eta_S/\eta_P$. The following list summarizes the important results of simulations detailed in Section \ref{sec:simulations}. 

\begin{enumerate}
    \item When astigmatism aberrations are present, applying aberrations in other Zernikes which are not opposite to initially present aberrations reach null coupling local minima
    \item Astigmatism and coma corrections are the amplitude of astigmatism to the one which minimizes null coupling
    \item Defocus and trefoil corrections are the amplitude to the one which maximizes mean coupling when scanning eight azimuthally distributed points at 60mas
\end{enumerate}

These conclusions inform the following general NCPA calibration procedure:

\begin{enumerate}
    \item Roughly center the coupling map by estimating the center of the ringlike shape of the radial maxima.
    \item Tune defocus by scanning eight equally distributed azimuthal points at 60mas with multiple amplitudes of applied defocus and apply the amplitude which maximizes the mean coupling.
    \item Tune astigmatism terms by scanning the null with multiple amplitudes of applied astigmatism and apply the amplitude of astigmatism which minimizes null coupling. Tune astigmatism orders 1-3 (Zernikes of Noll indices 5, 6, 12, 13, 23, and 24). It may be necessary to repeat this process twice to ensure that the presence of multiple orders of astigmatism error has not distorted the coupling effects.
    \item Tune coma by scanning the null with multiple amplitudes of applied coma and apply the amplitude which minimizes null coupling. Tune coma orders 1-2.
    \item Tune trefoil by scanning the eight equally distributed azimuthal points at 60mas to maximize the mean coupling. It is generally only necessary to include 1st order trefoil since trefoil aberrations have a weak effect on null depth and off-axis coupling.
    \item Recenter the coupling map with the deepest null point at the center.
    \item Rescan defocus in the same fashion as was done before.
    \item Scan the null and the eight off axis points to estimate a mean ratio $\eta_S/\eta_P$. If satisfactory, the process can be complete. If unsatisfactory, start the process again, focusing on astigmatism terms.
\end{enumerate}

\section{MODEL CALIBRATION}

\begin{figure}
\centering
  \begin{subfigure}[t]{.43\textwidth}
  \centering
    \includegraphics[width=.9\textwidth]{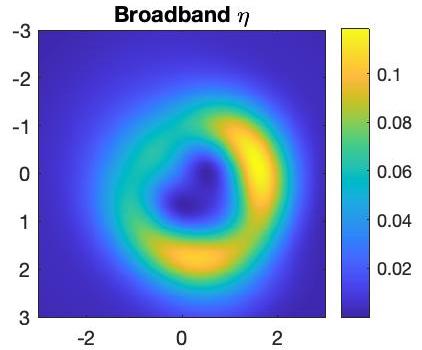}
    \caption{\label{fig:badstart}Aberrations in Zernikes 2-10 respectively (waves RMS): 0.1, 0.1, 0.05, 0.01, -0.02, 0.025, 0.025, 0.02, -0.01}
  \end{subfigure}
  \begin{subfigure}[t]{.43\textwidth}
  \centering
    \includegraphics[width=.9\textwidth]{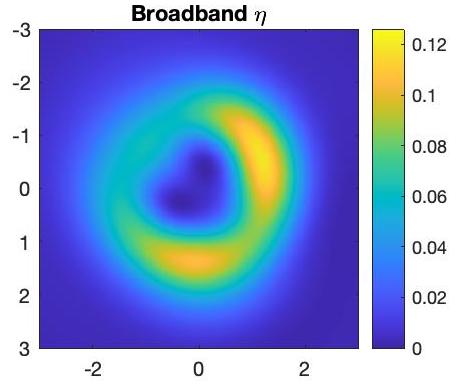}
    \caption{\label{fig:badcentered}After correcting tip/tilt}
  \end{subfigure}
  \newline
  \begin{subfigure}[t]{.43\textwidth}
  \centering
    \includegraphics[width=.9\textwidth]{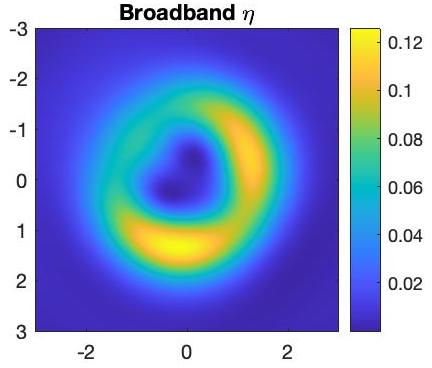}
    \caption{\label{fig:badfocused}After correcting defocus}
  \end{subfigure}
  \begin{subfigure}[t]{.43\textwidth}
  \centering
    \includegraphics[width=.9\textwidth]{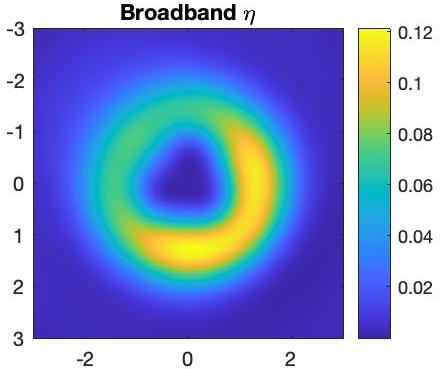}
    \caption{\label{fig:badnoastig}After correcting astigmatism}
  \end{subfigure}
  \newline
  \begin{subfigure}[t]{.43\textwidth}
  \centering
    \includegraphics[width=.9\textwidth]{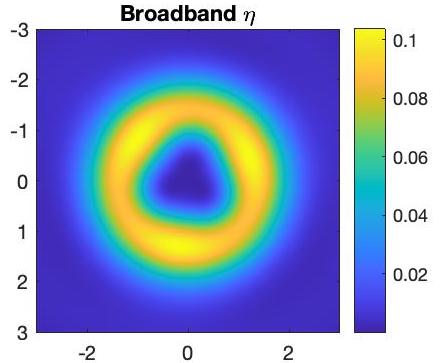}
    \caption{\label{fig:badnocoma}After correcting coma}
  \end{subfigure}
  \begin{subfigure}[t]{.42\textwidth}
  \centering
    \includegraphics[width=.9\textwidth]{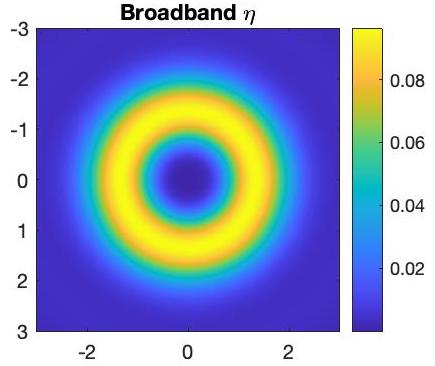}
    \caption{\label{fig:goodend}After correcting trefoil; no aberrations}
  \end{subfigure}
  \newline
\caption{\label{fig:startandend}Simulated coupling efficiency maps starting with an example starting condition map, the final map, and the intermediate maps resulting from the steps of the calibration procedure.}
\end{figure}

This section describes the results of using the eight step procedure listed above on the coupling map depicted in Figure \ref{fig:badstart}. This starting map is simulated and the initial aberrations are known so this section summarizes a proof of concept. It is clear that there are aberrations in Figure \ref{fig:badstart}, but the exact aberrations cannot be identified by eye so the calibration procedure will be necessary. For this process, the only aberrations present are first order Zernike polynomials. The initial aberrations are as follows for Zernikes 2-10 respectively (in waves RMS): 0.1, 0.1, 0.05, 0.01, -0.02, 0.025, 0.025, 0.02, -0.01.

Step 1 of the procedure is recentering the coupling map. Though null splitting is present, it is fairly clear that translating the coupling map to its position in Figure \ref{fig:badcentered} centers the donut shape. This translation corresponds to aberrations of -0.1 waves RMS in Zernikes 2 and 3 which are opposite to the initial aberrations of 0.1 waves RMS.

\begin{figure}
\centering
\includegraphics[width=0.8\textwidth]{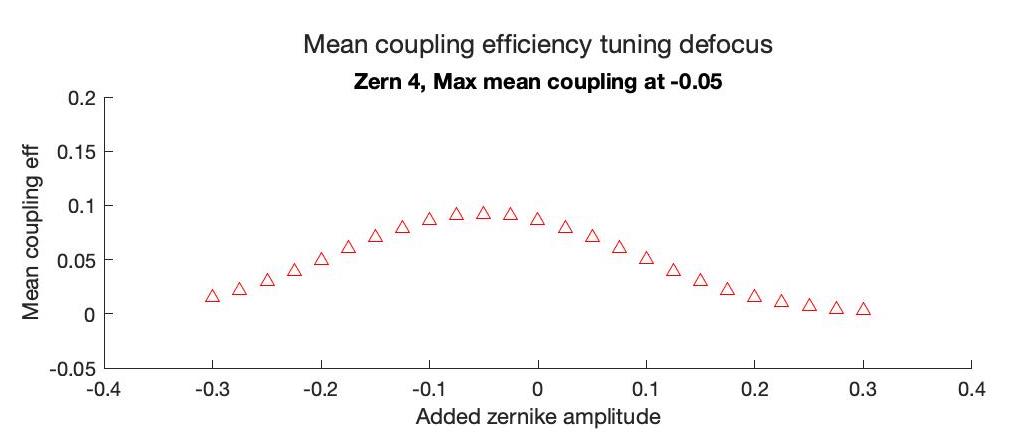}
\caption{\label{fig:focusing}Simulated mean coupling efficiency of eight azimuthally distributed points at 60mas with added defocus amplitudes ranging from -0.3 waves RMS to 0.3 waves RMS. There is an initial aberration of 0.05 waves RMS of defocus and other aberrations in astigmatism, coma, and trefoil. The maximum mean off axis coupling occurs when adding -0.05 waves RMS, which is opposite to the initial aberration.}
\end{figure}

Step 2 of the procedure is focusing the coupling map. As previously discussed, defocus decreases and spreads the radial maxima, so defocus can be corrected by applying the amplitude that maximizes off-axis coupling at 60mas. Figure \ref{fig:focusing} depicts the simulated mean off-axis coupling at eight azimuthal points at 60mas that occurs when different amplitudes of defocus are applied on at a time to the centered coupling map. As is shown, the maximum coupling occurs when -0.05 waves RMS of defocus is added, which are opposite to the initially added aberration of 0.05 waves RMS. After applying this correction, the coupling map looks like Figure \ref{fig:badfocused}. The visual difference between Figures \ref{fig:badcentered} and \ref{fig:badfocused} is subtle but the area of the light blue haze is smaller in the focused map than in the defocused map.

\begin{figure}
\centering
\includegraphics[width=0.8\textwidth]{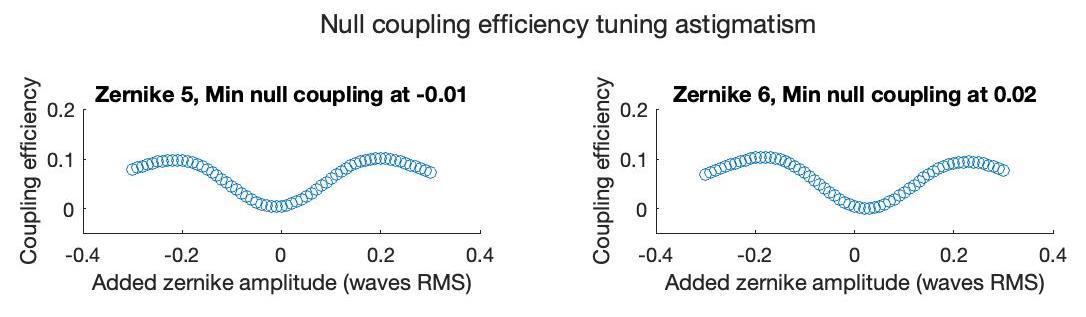}
\caption{\label{fig:removingastig}Simulated null coupling efficiency with added amplitudes of Zernikes 5 and 6 (first order astigmatism) ranging from -0.3 waves RMS to 0.3 waves RMS. There is an initial aberration of 0.01 waves RMS of Zernike 5, -0.02 waves RMS of Zernike 6 and other aberrations in coma and trefoil. The minimum null coupling occurs when adding -0.01 waves RMS of Zernike 5 or 0.02 waves RMS of Zernike 6, which are opposite to the initial aberrations.}
\end{figure}

Step 3 of the calibration procedure is tuning astigmatism by measuring null depth with different added amplitudes of astigmatism. Figure \ref{fig:removingastig} depicts the null depth that occurs when adding multiple different amplitudes of Zernikes 5 and 6 one at a time to the focused coupling map in Figure \ref{fig:badfocused}. The minimum null coupling occurs when adding -0.01 waves RMS of Zernike 5 and 0.02 waves RMS of Zernike 6 independently. These corrections are opposite to the initially added aberrations, which confirms that first order astigmatism aberrations can be detected in the presence of coma and trefoil aberrations as well as other first order astigmatism aberrations. Figure \ref{fig:badnoastig} depicts the resulting coupling map after applying the corrections to astigmatism. Though this map is still very asymmetric, the null has been unified again, which is to expected since astigmatism aberrations cause strong null splitting.

\begin{figure}
\centering
\includegraphics[width=0.8\textwidth]{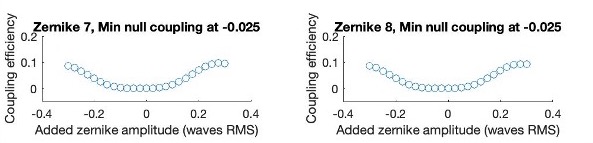}
\caption{\label{fig:removingcoma}Simulated mean coupling efficiency of eight azimuthally distributed points at 60mas with added amplitudes of Zernikes 7 and 8 ranging from -0.3 waves RMS to 0.3 waves RMS. There are initial aberrations of 0.025 waves RMS in Zernikes 7 and 8 and other aberrations in trefoil. The maximum mean off axis coupling occurs when adding -0.025 waves RMS in Zernikes 7 and 8 independently, which are opposite to the initial aberrations.}
\end{figure}

Step 4 of the procedure is tuning coma, which can be done using the same method as is used for astigmatism. Figure \ref{fig:removingcoma} depicts the null depth that occurs when adding multiple different amplitudes of Zernikes 7 and 8 one at a time to the coupling map in Figure \ref{fig:badnoastig}. The deepest nulls occur when adding -0.025 waves RMS of both Zernikes 7 and 8, which are opposite to the initially added aberrations. The resulting coupling map after applying these corrections is shown in Figure \ref{fig:badnocoma}.

\begin{figure}
\centering
\includegraphics[width=1\textwidth]{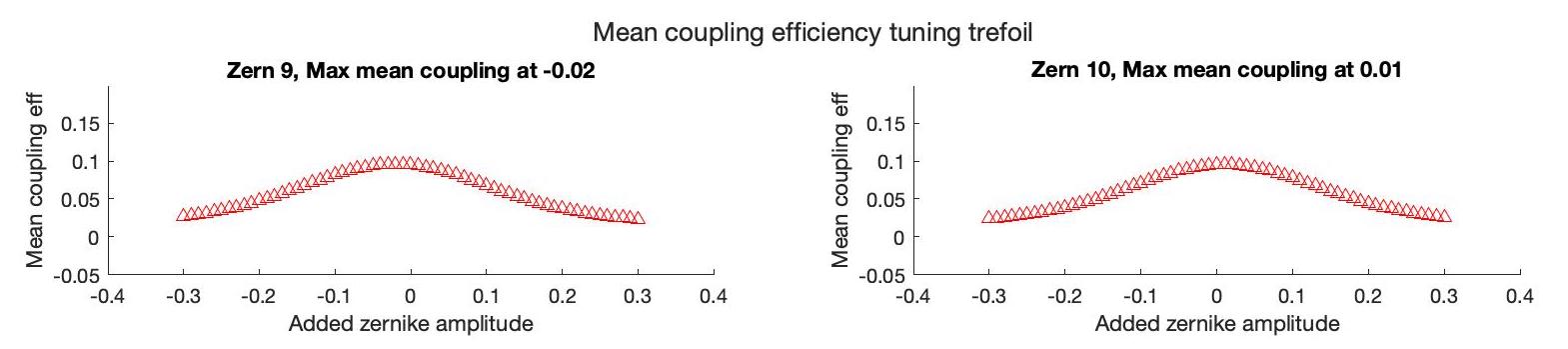}
\caption{\label{fig:tuningtrefoil}Simulated mean coupling efficiency of eight azimuthally distributed points at 60mas with added amplitudes of Zernikes 9 and 10 ranging from -0.3 waves RMS to 0.3 waves RMS. There are initial aberrations of 0.02 waves RMS in Zernike 9 and -0.01 in Zernke 10. The maximum mean off axis coupling occurs when adding -0.02 waves RMS of Zernike 9 or 0.01 waves RMS of Zernike 10, which are opposite to the initial aberrations.}
\end{figure}

The 5th step of the procedure is tuning trefoil, which is done in the same manner which was used for defocus. Figure \ref{fig:tuningtrefoil} depicts the simulated mean off-axis coupling at eight azimuthal points at 60mas that occurs when different amplitudes of Zernikes 9 and 10 are applied one at a time to the coupling map in Figure \ref{fig:badnocoma}. As is shown, the maximum coupling occurs when -0.02 waves RMS of Zernike 9 and 0.01 waves RMS of Zernike 10 are added, which are opposite of the initially added aberrations. After applying these corrections, the coupling map is the perfect donut which is shown in Figure \ref{fig:goodend}!

Because of the assumptions made for this example procedure, it was not necessary to go through steps 6 and 7, or to re-scan astigmatism a second time as suggested in step 8. The map was perfectly centered from the beginning so the corrections made were precisely opposite to the initially added corrections. Further, there were no initially added aberrations in higher orders of astigmatism so scanning the first order terms once accurately indicated the appropriate corrections. However, in the more realistic case where aberrations exist in higher orders of astigmatism, scanning all orders of astigmatism twice should provide a close estimate of the existing aberrations. Regardless, under the assumptions, the calibration procedure outlined in Section \ref{sec:method} works as expected.

\section{CONCLUSIONS AND FUTURE WORK}

The calibration procedure in Section \ref{sec:method} takes into account the important considerations for calibrating KPIC VFN. Notably, in a preliminary analysis of off-sky calibration performance, this method achieved a ratio $\eta_S/\eta_P\approx1/25$. Also, a high level of visual symmetry was achieved, although further data processing will be needed to quantify this result. The procedure in total took about 40 minutes, which is a reasonable compromise between efficiency and consideration of important effects.

\begin{figure}
\centering
  \begin{subfigure}[t]{1\textwidth}
  \centering
    \includegraphics[width=1\textwidth]{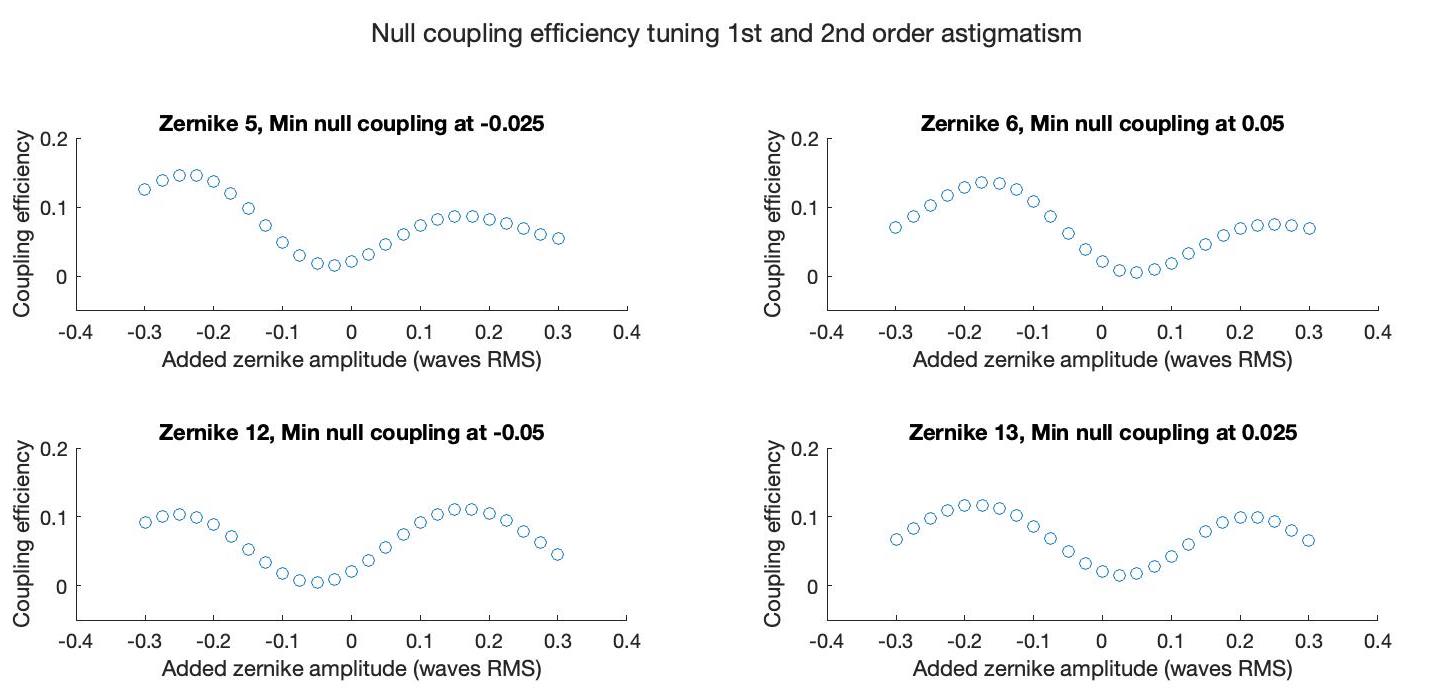}
    \caption{\label{fig:multiastig1}}
  \end{subfigure}
  \newline
  \begin{subfigure}[t]{1\textwidth}
  \centering
    \includegraphics[width=1\textwidth]{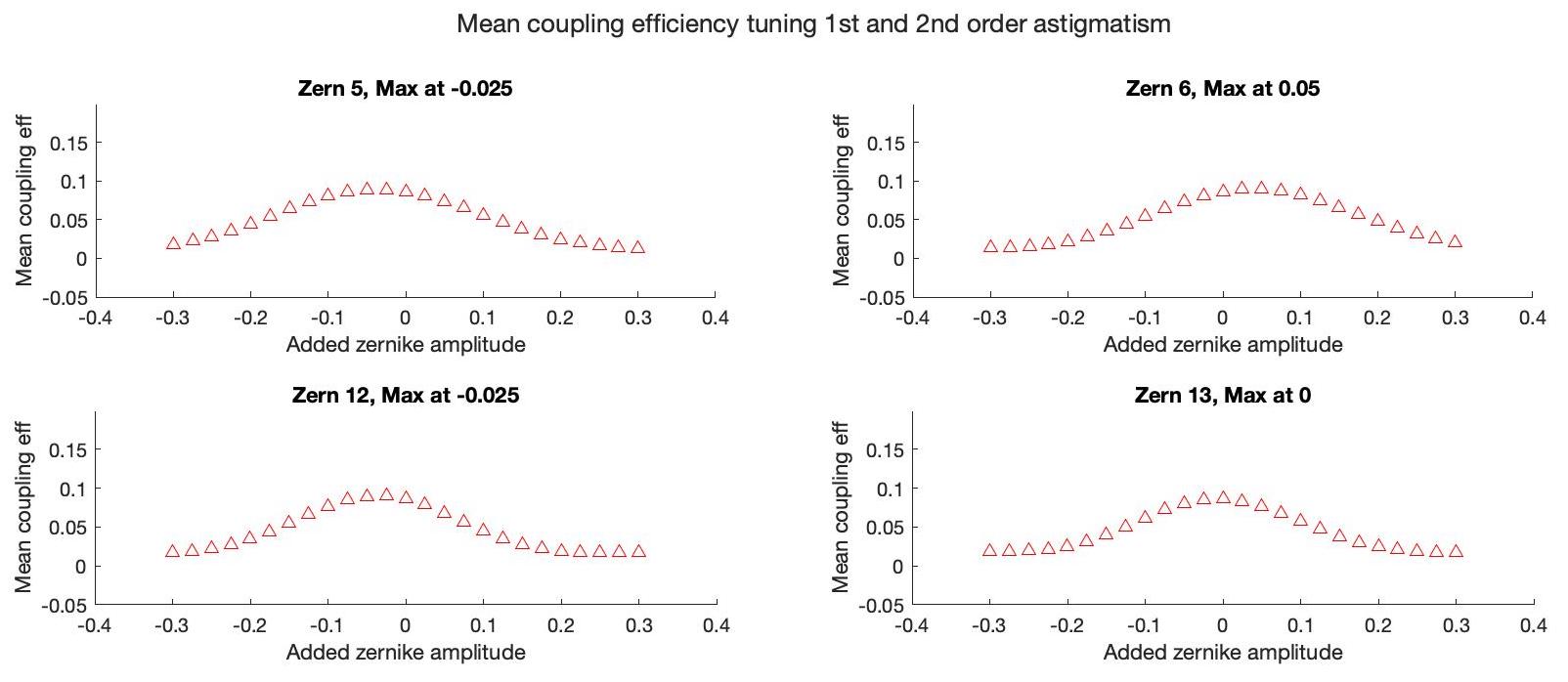}
    \caption{\label{fig:multiastig2}}
  \end{subfigure}
\caption{\label{fig:multiastig}Simulated null coupling efficiency and mean coupling efficiency of eight azimuthally distributed points at 60mas with added amplitudes of Zernikes 5, 6, 12, and 13 ranging from -0.3 waves RMS to 0.3 waves RMS. There are initial aberrations in Zernikes 5, 6, 12, and 13 respectively (in wave RMS): 0.05, -0.025, 0.025, and 0.025.}
\end{figure}

One potential issue in this method is that there is degeneracy in tuning the null when there are multiple orders of astigmatism aberrations present. Figure \ref{fig:multiastig1} depicts the null depth resulting from of adding different amplitudes of Zernikes 5, 6, 12, and 13 (first and second order astigmatism) one at a time to a coupling map which already has the following aberrations in Zernikes 5, 6, 12, and 13 respectively (in wave RMS): 0.05, -0.025, 0.025, and 0.025. Figure \ref{fig:multiastig2} depicts the mean off-axis coupling that occurs under the same conditions.

For both scanning the null and off-axis points, the added amplitude which either minimizes null coupling or maximizes off-axis coupling is (in this example) never opposite to the initially added amplitude. There are three possible solutions: to add more azimuthal points, to run a multidimensional scan with multiple Zernikes at once, or to cycle through detecting and correcting astigmatism aberrations multiple times. Adding more points is impractical as there will always be higher order astigmatism terms whose symmetry will not be resolved. Further, astigmatism has eight-fold azimuthal symmetry so it is doubtful that this would help. Next, as discussed previously, multidimensional scans are inefficient and should be avoided if possible. 

The remaining solution requires noting that, though the results are always incorrect, the amplitudes of Zernikes 5 and 6 which minimize the null and maximize off-axis coupling are in the correct direction of the appropriate corrections. The initially added aberrations were 0.05 waves RMS in Zernike 5 and -0.025 waves RMS in Zernike 6. The amplitudes which minimize null coupling (and also maximize off-axis coupling) are -0.025 in Zernike 5 and 0.05 in Zernike. Applying these corrections would result in a net aberration of 0.025 waves RMS in Zernike 5 and 0.025 in Zernike 6. So, while the corrections were in the correct direction of the appropriate corrections, the remaining aberrations are still considerable. Scanning and correcting one astigmatism term at a time and repeating each several times may eventually lead to the appropriate net correction. This is the goal of the current procedure. However, since null depth is so sensitive to astigmatism, it may be necessary to implement a multidimensional scan to ensure that as much astigmatism is removed as is possible. 

Performing a multidimensional scan for astigmatism would look like applying multiple combinations of Zernikes 5, 6, 12, and 13, as well as possibly Zernikes 23 and 24 (third order astigmatism) to determine which combination of corrections minimizes null coupling. Though this would take longer, it may significantly reduce null coupling. Next steps include simulating such a multidimensional scan followed by testing its performance on-sky.

Other next steps will involve processing on-sky data that has been obtained using the KPIC VFN mode and the calibration procedure listed here. This data will provide insight into the exact throughput ratio between null and off-axis points as well as the symmetry of the coupling map. With these results and planned next steps, this project will provide a basis for the important considerations for VFN calibrations at KPIC and beyond.




 
\clearpage
\bibliography{report} 
\bibliographystyle{spiebib} 

\end{document}